\documentclass[aps,twocolumn,preprintnumbers,amsmath,amssymb,nofootinbib,superscriptaddress,notitlepage]{revtex4-1}

\usepackage{epsfig}

\begin{document}

\title{Are there near-threshold Coulomb-like Baryonia?}

	\author{Li-Sheng Geng}\email{lisheng.geng@buaa.edu.cn}
	\author{Jun-Xu Lu} 
	\author{M. Pavon Valderrama}\email{mpavon@buaa.edu.cn}
\affiliation{School of Physics and Nuclear Energy Engineering, \\
International Research Center for Nuclei and Particles in the Cosmos and \\
Beijing Key Laboratory of Advanced Nuclear Materials and Physics, \\
Beihang University, Beijing 100191, China} 
	\author{Xiu-Lei Ren}
\affiliation{State  Key  Laboratory  of  Nuclear  Physics  and  Technology,
School  of  Physics,  Peking  University,  Beijing  100871,  China}
\affiliation{
Institut f\"ur Theoretische Physik II, Ruhr-Universit\"at  Bochum,
D-44780 Bochum, Germany}

\date{\today}


\begin{abstract} 
\rule{0ex}{3ex}
The $\Lambda_c(2590) \Sigma_c$ system can exchange a pion near
the mass-shell. 
Owing to the opposite intrinsic parity of
the $\Lambda_c(2590)$ and $\Sigma_c$, the pion is exchanged in S-wave.
This gives rise to a Coulomb-like force that might be able
to bind the system.
If one takes into account that the pion is not exactly on the mass shell,
there is a shallow S-wave state,
which we generically call the $Y_{cc}(5045)$ and $Y_{c\bar c}(5045)$ 
for the $\Lambda_c(2590) \Sigma_c$ and $\Lambda_c(2590) \bar{\Sigma}_c$
systems respectively.
For the baryon-antibaryon case this Coulomb-like force is independent of spin:
the $Y_{c\bar c}(5045)$ baryonia will appear either
in the spin $S=0$ or $S=1$ configurations
with G-parities $G=(-1)^{L+S+1}$.
For the baryon-baryon case the Coulomb-like force is attractive
in the spin $S=0$ configuration,
for which a doubly charmed molecule is expected to form near the threshold.
This type of spectrum might be very well realized in other molecular states
composed of two opposite parity hadrons with the same spin and
a mass difference close to that of a pseudo-Goldstone boson,
of which a few examples include the $\Lambda(1405) N$, $\Lambda(1520) \Sigma^*$,
$\Xi(1690) \Sigma$, $D_{s0}^*(2317) D$ and $D_{s1}^*(2460) D^*$ molecules.
\end{abstract}

\maketitle

The discovery of the $X(3872)$ a decade ago by Belle~\cite{Choi:2003ue}
opened a new era in hadron spectroscopy.
The $X(3872)$ was the first member of a growing family of states
above the open charm and bottom thresholds that do not fit
into the traditional charmonia and bottomonia spectra,
the XYZ states.
Their seemingly unending variety requires a series of explanations which
includes
tetraquarks~\cite{Maiani:2004vq,Ebert:2005nc,Stancu:2009ka,Wang:2013vex,Maiani:2014aja},
pentaquarks~\cite{Yuan:2012wz,Maiani:2015vwa,Lebed:2015tna,Li:2015gta,Wang:2015epa},
hybrids~\cite{Zhu:2005hp,Kou:2005gt}
threshold effects~\cite{Bugg:2004rk,Chen:2011pu,Chen:2011pv,Guo:2015umn},
hadrocharmonia~\cite{Dubynskiy:2008mq},
baryocharmionia~\cite{Kubarovsky:2015aaa}
and of course molecules~\cite{Voloshin:1976ap,DeRujula:1976qd,Thomas:2008ja,Ding:2008gr,Lee:2009hy,Kang:2016jxw}
(see Refs.~\cite{Brambilla:2010cs,Faccini:2012pj,Guo:2017jvc} for reviews).
In fact the most promising explanation for the $X(3872)$ is
that of a molecular state,
i.e. a bound state of two hadrons, in particular the $D^{0}\bar{D}^{0*}$.
Yet the $X(3872)$ is not the only instance.
Other molecular candidates include the $Z_c$'s
in the charm sector~\cite{Ablikim:2013mio,Liu:2013dau}
and the $Z_b$'s in the bottom one~\cite{Belle:2011aa,Adachi:2012im}.
The recent discovery of the $P_c(4380)^+$ and $P_c(4450)^+$
pentaquarks~\cite{Aaij:2015tga}
might add the later to the family of molecular candidates:
the $P_c(4450)^+$ could be a $\Sigma_c \bar{D}^*$ or $\Sigma_c^* \bar{D}^*$
molecule~\cite{Chen:2015loa,Chen:2015moa,Roca:2015dva}.
Hadron molecules were theorized three decades ago in analogy
with the deuteron~\cite{Voloshin:1976ap,DeRujula:1976qd}:
the exchange of light mesons generates a force that
might very well be able to bind a hadron system.
Here we will consider a particular type of hadron molecule
--- the $\Lambda_{c1} \Sigma_c$ baryon-baryon and
$\Lambda_{c1} \bar{\Sigma}_c$ baryon-antibaryon systems ---
where the exchange of a pion near the mass shell mimics
the time-honored Coulomb-potential.

In a recent work~\cite{Geng:2017hxc}
we have discussed a molecular explanation for the $P_c(4450)^+$
which besides the usual $\Sigma_c \bar{D}^*$ also involves a
$\Lambda_{c1} \bar{D}$ component,
where the $\Lambda_{c1}$ denotes the $\Lambda_c(2590)$.
This generates a {\it vector force} --- the equivalent of a tensor force
but with angular momentum $L=1$ instead of $L=2$ --- that might play
an important role in binding and might trigger
discrete scale invariance if strong enough.
A curious thing happens if the $\bar{D}^*$ and $\bar{D}$ piece of
this molecule is changed by a $\bar{\Lambda}_{c1}$ and $\bar{\Sigma}_{c}$:
we obtain a Coulomb-like $1/r$ potential.
This opens the prospect of a molecule exhibiting a hydrogen-like spectrum,
which will be broken at low energies owing
to the off-shellness of the pion.
Besides we find it indeed remarkable that there is the possibility of making
relatively concrete predictions for heavy hadron molecules
without a strong requirement of guessing the short-range physics
or using arbitrary form-factors~\footnote{The standard tensor force diverges
as $1/r^3$ at short distances. If attractive, the eigenvalues of
the hamiltonian do not have a lower bound, i.e. the energy of
the fundamental state goes to minus infinity. The solution
is to include a form factor.
This does not happen with a Coulomb or Yukawa potential.
}.
Yet we will use these type of assumptions to check the robustness
of the results.
 
The mechanism by which a $\Lambda_{c1} \Sigma_c$ molecule
binds is generic and applies to other hadron systems.
The exchange of a pseudo-Goldstone boson between a negative and positive parity
hadron pair with the same spin leads to a Yukawa potential.
In addition there is a tendency for negative (positive) parity hadrons
to be near the threshold of a positive (negative) parity hadron and
a pseudo-Goldstone boson.
As a consequence the range of the interaction will be considerably
larger than expected.
In the light sector examples are
the $\Lambda(1405)$-$N \bar{K}$~\cite{Magas:2005vu,Hyodo:2011ur}
and $\Lambda(1520)$-$\Sigma^* \pi$~\cite{Aceti:2014wka}.
The $\Xi(1690)$-$\Sigma \bar{K}$ system might be a third instance
if the quantum numbers of the $\Xi(1690)$
turn out to be $\frac{1}{2}^{-}$, as suggested in~\cite{Aubert:2008ty}.
In the heavy sector, if we restrict ourselves to experimentally known hadrons,
besides $\Lambda_{c1}$-$\Sigma_c \pi$, we also have
$D_{s0}^*(2317)$-$D K$ and $D_{s1}^*(2460)$-$D^* K$~\cite{Guo:2006fu,Guo:2006rp,Guo:2011dd} (where a recent calculation shows that
the $D_{s0}^* D$ and $D_{s1}^* D^*$ systems are
bound~\cite{SanchezSanchez:2017xtl}). 
Theoretical explorations~\cite{GarciaRecio:2012db,Lu:2014ina,Garcia-Recio:2015jsa}
indicate that this type of heavy hadron pairs is common.

We explain now the Coulomb-like potential
for the $Y_{cc}$ ($\Lambda_{c1} \Sigma_c$) and
$Y_{c \bar c}$ ($\Lambda_{c1} \bar{\Sigma}_c$) molecules.
We begin with the baryon-antibaryon case,
for which we consider states in the isospin basis with well-defined G-parity
\begin{eqnarray}
| \Lambda_{c1} \bar{\Sigma}_c(\eta) \rangle =
\frac{1}{\sqrt{2}}
\left[
| \Lambda_{c1} \bar{\Sigma}_c \rangle +
\eta\, | \Sigma_c \bar{\Lambda}_{c1} \rangle
\right] \, ,
\end{eqnarray}
where $G = \eta (-1)^{L+S}$,
for which the one pion exchange (OPE) potential reads
\begin{eqnarray}
V_{\rm OPE}(r) =
\eta\,\frac{h_2^2 \omega_{\pi}^2}{4 \pi f_{\pi}^2}\frac{e^{-\mu_{\pi} r}}{r} \, , 
\label{eq:OPE}
\end{eqnarray}
where $\mu_{\pi}^2 = m_{\pi}^2 - \omega_{\pi}^2$,
with $\omega_{\pi} = m_{\Lambda_{c1}} - m_{\Sigma_c}$.
The potential is derived from the Lagrangian of Cho~\cite{Cho:1994vg}
and the formalism of Refs.~\cite{Valderrama:2012jv,Lu:2017dvm},
which leads to the non-relativistic amplitude
$\mathcal{A}(\Lambda_{c1} \to \Sigma_c \pi) = h_2 \, \omega_{\pi} / f_{\pi}$.
We take $f_{\pi} = 130\,{\rm MeV}$ and
$h_2 = 0.63 \pm 0.07$~\cite{Cheng:2015naa}
(this value is based on a theoretical analysis of the $\Lambda_{c1}$ decays
and is almost identical to $h_2 = 0.60 \pm 0.07$
from CDF~\cite{Aaltonen:2011sf}).
In the isospin symmetric limit $\mu_{\pi}^2 < 0$, yielding a complex potential
similar to the one in the $X(3872)$ (except that it is much stronger).
However if we consider the isospin components of the $Y_{c\bar{c}}(5045)$
\begin{eqnarray}
| 1 \, , +1 \rangle_I &=& \frac{1}{\sqrt{2}} \, \left[
| \Lambda_{c1}^{+} \bar{\Sigma}^0_c \rangle  + \eta\,
| \Sigma_{c}^{++} \bar{\Lambda}^-_{c1} \rangle \right] \, , \\
| 1 \, , \phantom{+} 0 \rangle_I &=& \frac{1}{\sqrt{2}} \, \left[
- | \Lambda_{c1}^{+} \bar{\Sigma}^-_c \rangle  + \eta\,
| \Sigma_{c}^{+} \bar{\Lambda}^-_{c1} \rangle \right] \, , \\
| 1 \, , \, - 1 \rangle_I &=& \frac{1}{\sqrt{2}} \, \left[
| \Lambda_{c1}^{+} \bar{\Sigma}^{--}_c \rangle  + \eta\,
| \Sigma_{c}^{0} \bar{\Lambda}^-_{c1} \rangle \right] \, ,
\end{eqnarray}
upon closer inspection we realize that the $m_I = \pm 1$
states exchange a charged pion and the $m_I = 0$ a neutral pion.
For the charged pion case $\mu_{\pi^{\pm}}^2 > 0$ and
the OPE potential displays exponential decay at long distances.
The effective pion mass is $\mu_{\pi^{\pm}} \simeq 18\,{\rm MeV}$,
which translates into eight times the standard range of OPE.
If we consider the reduced potential instead,
we can define the equivalent of the Bohr radius as
\begin{eqnarray}
2\mu_{Y} V(r) &=& \eta \frac{2}{a_B}\,
\frac{e^{-\mu_{\pi} r}}{r} \, ,
\end{eqnarray}
where $\mu_Y$ is the reduced mass and $a_B$ is given by
\begin{eqnarray}
a_B &=&
\frac{4 \pi\,f_{\pi}^2}
{\mu_{Y}\,h_2^2 \, \omega_{\pi}^2}
= 4.4^{+1.2}_{-0.8}\,{\rm fm} \, ,
\end{eqnarray}
where for the masses of the $\Lambda_{c1}$ and $\Sigma_c$ we take the values
of the PDG~\cite{Olive:2016xmw}.
It is also interesting to consider the Bohr momentum
\begin{eqnarray}
\label{eq:bohr-momentum}
\gamma_B = \frac{1}{a_B} = 45^{+10}_{-10}\,{\rm MeV} \, .
\end{eqnarray}
If the $\Lambda_{c1}$ and $\Sigma_c$ were stable and the pion were
on the mass shell, for $\eta=-1$ and in the absence of short range
forces we will have the Coulomb-like spectrum
\begin{eqnarray}
E_{n,l} = - \frac{1}{2 \mu_{Y}}\,
{\left( \frac{\gamma_B}{n+l+1} \right)}^2 \, ,
\end{eqnarray}
with $l$ the angular momentum and where we take $n=0$
for the ground state in each partial wave.

However neither the pion is on the mass shell
nor the $\Lambda_{c1}$ and $\Sigma_c$ are stable.
The finite effective mass of the pion means that
Coulomb-like bound state are expected to survive
only if their binding momentum fulfills the condition
\begin{eqnarray}
\frac{\gamma_B}{n+l+1} > \mu_{\pi} \, ,
\end{eqnarray}
which can only be met for $n + l + 1 \leq 2$ at best,
leaving room for few bound states at most.
In fact concrete calculations show that only
the $n=0$ S-wave states survives.
It is also important to consider the possible impact of the finite width of
the $\Lambda_{c1}$ and $\Sigma_{c}$ heavy baryons, about $2\,{\rm MeV}$.
In the absence of short-range forces the energy of
the fundamental state is expected to be
$E = -0.8^{+0.4}_{-0.3}\,{\rm MeV}$,
which is about half the width of the heavy baryons.
The authors of Ref.~\cite{Guo:2011dd} argue that the width of the components
can be ignored if their lifetime is ample enough for the formation of
the bound state.
This is equivalent to the condition $\Gamma \ll m$
with $m$ the mass of the exchanged meson.
If instead of the physical pion mass
we take the effective pion mass $\mu_{\pi} \sim 20\,{\rm MeV}$
to be on the safe side, there is still plenty of time
for the formation of the bound state before its components decay.
Hence we expect it to survive.

\begin{figure}[ttt]
\begin{center}
\includegraphics[width=9.5cm]{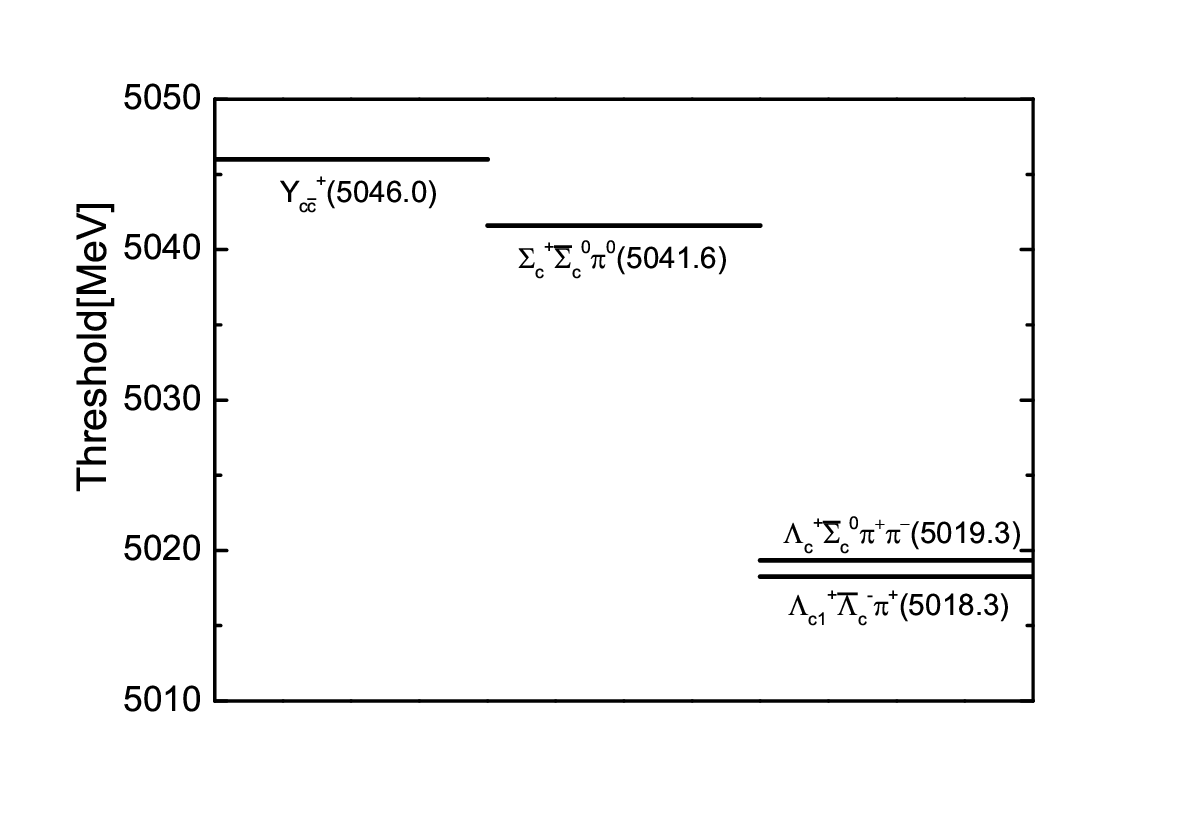}
\end{center}
\caption{
Location of the $Y_{c\bar{c}}^+$ ($\Lambda_{c1}^+ \bar{\Sigma}_c^0$) threshold
and its closest decay channels, the $\Sigma_c^+ \bar{\Sigma}_c^0 \pi^0$
($4.4\,{\rm MeV}$ below the threshold)
and the $\Lambda_{c1}^+ \to \Lambda_c \pi^+ \pi^-$ 
and $\Sigma^{++}_c \to \Lambda^+_c \pi^+$
($26.7$ and  $27.7\,{\rm MeV}$ below respectively). 
}
\label{Ycc-decays}
\end{figure}

Yet there is a second argument that calls for relative caution.
In a more complete analysis of bound states with unstable constituents,
Hanhart et al.~\cite{Hanhart:2010wh} proposed to check the dimensionless ratio
\begin{eqnarray}
\lambda = \frac{\Gamma_R}{2 E_R} \, ,
\end{eqnarray}
where $R$ refers to a decay channel of one of the constituents,
$\Gamma_R$ the partial decay width and $E_R$ the energy gap
to that decay channel.
The $Y_{c \bar c}$ decays induced by its components can be seen
in Fig.~\ref{Ycc-decays}.
If we consider the decays mediated by $\Sigma^{++}_c \to \Lambda^+_c \pi^+$
and $\Lambda_{c1}^+ \to \Lambda_c \pi^+ \pi^-$
(i.e. $Y_{c \bar c}^+ \to \Lambda_{c1}^{+} \bar{\Lambda}_c^{-} \pi^+$
and $Y_{c \bar c}^+ \to \Lambda_{c}^+ \bar{\Sigma}_c^0 \pi^+ \pi^-$)
we have $\lambda = 0.034$ and $\lambda = 0.045$ respectively.
The problem arises with the decay
$Y_{c \bar c}^{+} \to
(\Sigma_c^+ \bar{\Sigma}_c^0 + \Sigma_c^{++} \Sigma_{c}^{-}) \pi^0$.
It happens merely $4.37 \pm 0.49\,{\rm MeV}$ below
the $\Lambda_{c1}^+ \bar{\Sigma}_c^0$ threshold,
yet the $\Lambda_{c1}^+ \to \Sigma_c^+ \pi^0$ decay width,
though unknown, is comparable to that value.
From heavy baryon chiral perturbation theory~\cite{Cho:1994vg}
we expect this decay to be
\begin{eqnarray}
\Gamma(\Lambda_{c1}^+ \to \Sigma_c^+ \pi^0) = \frac{1}{2 \pi}\,
\frac{m_{\Sigma_c}}{m_{\Lambda_{c1}}}\,\frac{h_2^2 \omega_{\pi}^2}{f_{\pi}^2}\,
q_{\pi} \, , \label{eq:decay}
\end{eqnarray}
with $q_{\pi}$ the momentum of the outgoing pion,
leading to $2.1^{+0.5}_{-0.4}\,{\rm MeV}$
with the numbers we are using here.
For a molecular state near threshold,
this leaves the range $\lambda \sim 0.18-0.30$
with the central value $\lambda = 0.24$.
These numbers are not bad but not ideal either.
They can imply a sizeable distortion of the lineshapes~\cite{Hanhart:2010wh},
indicating that a more complete analysis of shallow
Coulomb-like baryonium states might be useful.
There is also the possibility that the interplay between
the $\Lambda_{c1} \to \Sigma_c \pi$ and $\Lambda_{c1} \to \Lambda_c \pi \pi$
will play in favor of neglecting the widths
in the shallow states.
For comparison purposes, previous speculations about the $Y(4260)$
as a $D_0 \bar{D^*}$/$D_1(2420) \bar{D}$ bound state owing
to a $e^{i | \mu_{\pi}| r}/r$ OPE potential~\cite{Close:2009ag}
(notice the complex exponential) are probably theoretically unsound
because of the large width of the P-wave heavy mesons~\cite{Filin:2010se},
where in this example the dimensionless parameter $\lambda_R$ is close to one.
A more thorough answer probably requires a full calculation in the line of
the one in Ref.~\cite{Baru:2011rs} for the $X(3872)$,
which included the $D\bar{D}^*$, $D^*\bar{D}$ and $D\bar{D}\pi$ channels.
The equivalent calculation for the $Y_{c\bar c}(5045)$ baryonium will
require the inclusion of the $\Lambda_{c1} \bar{\Sigma}_c$,
$\Sigma_{c} \bar{\Lambda}_{c1}$, $\Sigma_c \bar{\Sigma}_c$
and $\Sigma_c \bar{\Sigma}_c \pi$ channels (the last one being
also interesting because of the contribution of $\Sigma_c \pi$
to the $\Lambda_{c1}$
wave function~\cite{Long:2015pua,Long:2016oog,Guo:2016wpy}).
There is also the observation that the decay channels
$\Sigma_c^+ \bar{\Sigma}_c^0 \pi^{0}$ and $\Sigma_c^{++} \Sigma_{c}^{-} \pi^0$
do not appear as intermediate states in the $Y_{c \bar c}^+$.
This situation is different than in the $X(3872)$,
where the $D^0\bar{D}^0\pi^0$ can be a transient state
when the neutral pion is in flight,
which also happens in the $Y_{c\bar{c}}^0$
with the $\Lambda_{c1}^{+} \to \Sigma_c^{+} \pi^0$.
This is the reason why the OPE potential in the $Y_{c\bar{c}}^+$ is real,
while in the $X(3872)$ or in the $Y_{c\bar{c}}^0$ it acquires a complex part.
As a consequence the contribution from the aforementioned decay channels
might be important for the location of the $Y_{c\bar{c}}^0$
but not necessarily for the $Y_{c\bar{c}}^+$.

In the previous discussion we have only considered the states with $\eta = -1$,
for which the OPE potential is attractive.
For $\eta = +1$, though OPE is repulsive, there is the interesting feature
that the decay $Y_{c \bar c} \to \Sigma_c \bar{\Sigma}_c \pi$
is forbidden by C- and G-parity for an S-wave pion.
In fact the G-parity of the initial $Y_{c \bar c}$ is $G = (-1)^S$,
the final $\Sigma_c \bar{\Sigma}_c$ has $G = (-1)^S$ and
the pion has $G = -1$.
Vector meson exchange suggest that the short range interaction
between the $\Lambda_{c1}$ and $\bar{\Sigma}_c$ is repulsive,
but two-pion exchange is likely to be attractive
at intermediate distances.
If strong enough it could give rise to a bound state or a resonance above
the threshold owing to the Coulomb-like potential barrier
(which at $1\,{\rm fm}$ rises to $6$-$7$ $\rm MeV$).
Despite unlikely this type of baryonium is indeed very interesting
as the main decay mechanisms are forbidden and
hence we can expect a state narrower
than its components.

The $Y_{c \bar c}$ can also decay into charmonium /
charmed meson-antimeson pairs and light hadrons.
The intermediate states are hundreds of MeV
below the $\Lambda_{c1} \Sigma_c$ threshold
but might still affect the location of the bound states,
as happens for instance
in the nucleon-antinucleon case~\cite{Myhrer:1976by,Myhrer:1976ka}.
The mechanism behind these decays is of a short-range nature and
can be modelled with a complex potential,
for which we will present exploratory calculations later
(in line with the calculations of Ref.~\cite{Myhrer:1976ka}).
We will find that if binding is due to Coulomb-like OPE
then the $Y_{c \bar c}$ will be relatively unaffected by annihilation.

Finally we consider the baryon-baryon case,
for which the OPE potential reads 
\begin{eqnarray}
V_{\rm OPE}(r) =
- \, \frac{h_2^2 \omega_{\pi}^2}{4 \pi f_{\pi}^2}\frac{e^{-\mu_{\pi} r}}{r} 
\begin{pmatrix}
0 & 1 \\
1 & 0 
\end{pmatrix}
\, , 
\end{eqnarray}
where $\mu_{\pi}^2 = m_{\pi}^2 - \omega_{\pi}^2$
with $\omega_{\pi} = m_{\Lambda_{c1}} - m_{\Sigma_c}$.
Channels $1$ and $2$ are $\Lambda_{c1} \Sigma_c$ and
$\Sigma_c \Lambda_{c1}$ respectively.
The OPE potential is attractive for $S=0$,
for which the baryons take the configuration
\begin{eqnarray}
| Y_{cc} \rangle =
\frac{1}{\sqrt{2}}\,
\{
| \Lambda_{c1} \Sigma_{c} \rangle + | \Sigma_{c} \Lambda_{c1} \rangle 
\} \, , 
\end{eqnarray}
while for $S=1$ it happens to be repulsive~\footnote{The easiest way to see
  this is the formulation of extended Fermi-Dirac statistics
  for the $\Lambda_{c1}$ and $\Sigma_c$ baryons (in analogy to
  the neutron-proton case). In this case the wave function is antisymmetric
  under the exchange of particles $1$ and $2$, which means that
  the configuration that gives an attractive potential must have $S=0$.
  Another way to determine the sign is to notice that
  the potential is defined
  in the $\Lambda_{c1} \Sigma_c \to \Lambda_{c1} \Sigma_c$ channel, which implies
  the exchange of the $\Lambda_{c1}$ and $\Sigma_c$ fields
  in the final state, leading to a $(-1)^S$ overall factor
  (as in the $\Lambda(1405) N$ system~\cite{Uchino:2011jt}).
}.
The $S=0$ ($S=1$) long-range potential is indeed identical to that of
the baryon-antibaryon case with $\eta = -1$ ($\eta = +1$),
except that in the later case OPE is independent of spin.
The short-range physics is expected to be different
in the baryon-baryon and baryon-antibaryon systems.

\begin{figure}[hhh]
\begin{center}
\includegraphics[height=6.0cm]{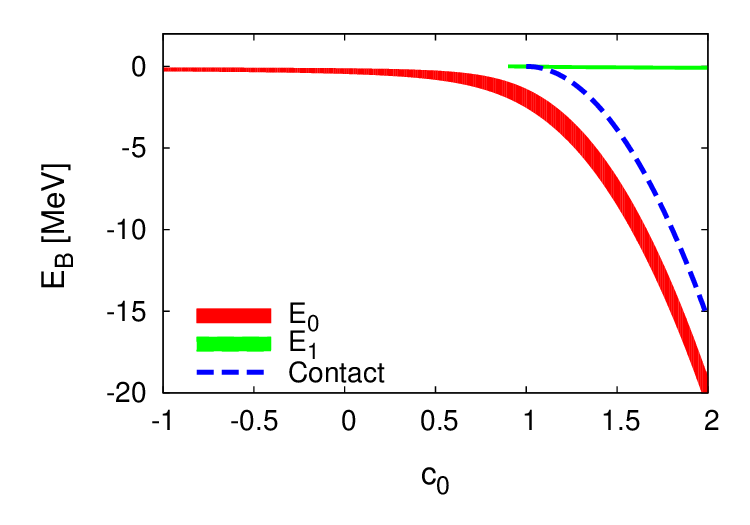}
\end{center}
\caption{
Binding energy for the $Y_{cc}$ and $Y_{c\bar c}$ states depending
on the relative strength of the short-range attraction.
In the absence of OPE or other long-range forces, $c_0 = 1$
corresponds to a bound state at threshold.
The blue-dashed line is the binding energy that results from
the short-range force alone.
The red band is the binding energy when one includes Coulomb-like OPE,
where the spread comes from the uncertainty in $h_2$.
The binding  stabilizes to $E_B = - 0.09^{+0.06}_{-0.08}\,{\rm MeV}$
for $c_0 \to -\infty$, i.e. for a hard-core at the cut-off radius. 
The green band is the energy of the second bound states,
which appears at $c_0 > 0.9^{+0.2}_{-0.4}$.
}
\label{Ycc-binding}
\end{figure}


We can quantify the discussion about the existence of the bound states
in the following way.
We will assume that the OPE potential is only valid above a certain cut-off
radius $R_c$, below which the interaction is described by a delta-shell
\begin{eqnarray}
V(r) = V_{\rm OPE}(r)\,\theta(r - R_c) + 
\frac{C_0}{4 \pi R_c^2}\,\delta(r - R_c) \, .
\label{eq:pot-full}
\end{eqnarray}
For convenience, instead of using the standard coupling $C_0$ we will
define the reduced coupling
\begin{eqnarray}
c_0 = -\frac{2 \mu_Y C_0}{4 \pi R_c} \, ,
\end{eqnarray}
where we have flipped the sign of the coupling such that the delta-shell
generates a bound state for $c_0 \geq 1$ with binding momentum
$\gamma = (c_0 - 1) / R_c$.
We can compute the spectrum as a function of $c_0$ and $R_c$.
If we choose $R_c = 1\,{\rm fm}$, which seems a sensible value,
we obtain the binding energies of Fig.~(\ref{Ycc-binding}).
For $c_0 \to - \infty$, which corresponds to a hard core at $R_c$,
the ground state survives with $E_B = - 0.09^{+0.06}_{-0.08}\,{\rm MeV}$.
For $c_0 = 1$, OPE shifts the binding energy from zero to
$E_B = -1.9^{+0.5}_{-0.6}$, which indicates a moderate contribution
from Coulomb-like OPE.
In addition for $c_0 > 0.9^{+0.2}_{-0.4}$ a shallow excited state appears.
We do not know the form of the short-range interaction between heavy baryons.
Phenomenological arguments (vector meson exchange) suggest
that for the $Y_{cc}$ ($Y_{c \bar c}$) short-range repulsion
(attraction) is more likely~\footnote{
  $\omega$ exchange will generate a central force that is repulsive
  (attractive) for $\Sigma_c \Lambda_{c1}$ ($\bar{\Sigma}_c \Lambda_{c1}$),
  while $\rho$ exchange will generate a non-diagonal spin-spin force
  which for $S=1$ ($\vec{\sigma}_1 \cdot \vec{\sigma}_2 = 1$)
  has the same (opposite) sign of OPE
  for $\Sigma_c \Lambda_{c1}$ ($\bar{\Sigma}_c \Lambda_{c1}$).
  The strength of $\omega$ exchange can be deduced from $SU(3)$-flavour
  symmetry and the OZI rule. The $\rho$ coupling can be deduced from
  vector meson dominance and the electromagnetic decay
  $\Gamma(\Lambda_{c1} \to \Sigma_c \gamma)$. Unfortunately the size of
  this partial width is unknown.
}.
However the cut-off radius $R_c = 1\,{\rm fm}$ probably lies
in an intermediate zone dominated by two-pion exchange
and other contributions which might be attractive.
Hence we expect the fundamental state of the double charmed $Y_{c c}$
(and maybe the $Y_{c \bar c}$ baryonium too)
to be deeper than the predictions from OPE alone
(thus implying the existence of a shallow excited state), though there is no
model-independent way to estimate
how much exactly.
Yet regardless of the short-range details a shallow state should survive.

\begin{figure}[ttt]
\begin{center}
  \includegraphics[height=4.2cm]{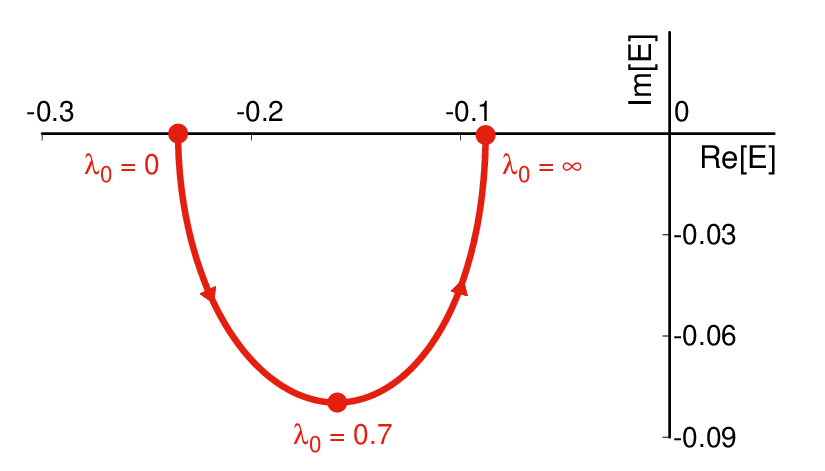}
  \includegraphics[height=4.2cm]{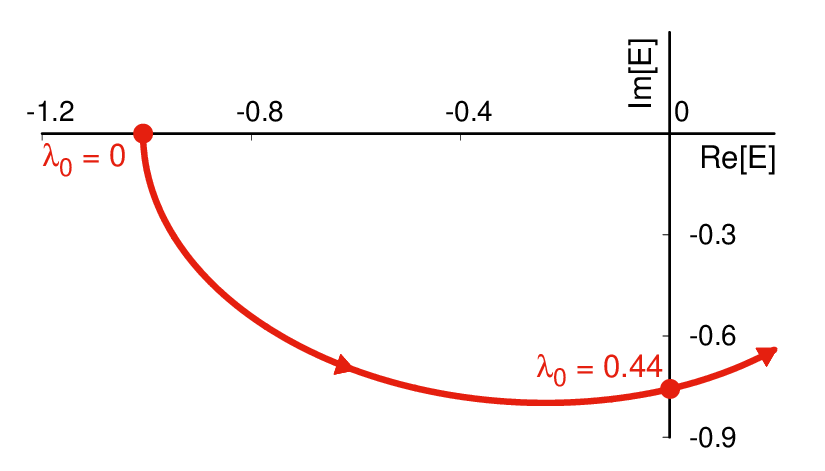}
  \includegraphics[height=4.2cm]{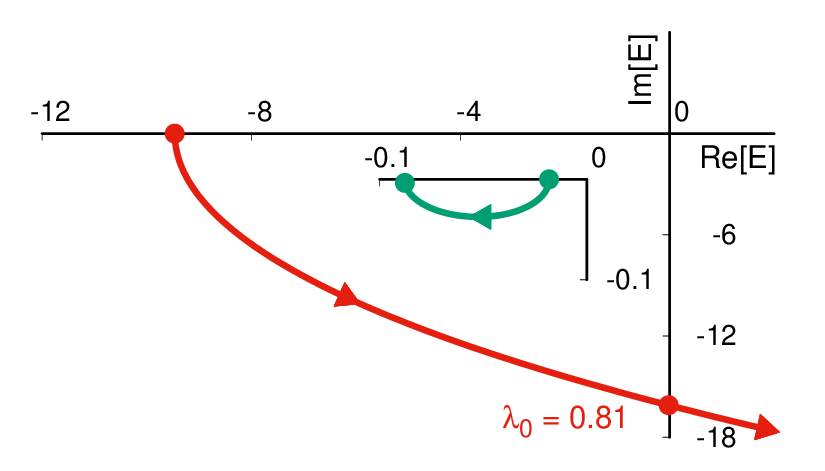}
\end{center}
\caption{
  Movement of the $Y_{c\bar c}$ pole in the complex plane
  when we take into account baryon-antibaryon annihilation (in MeV).
  This is done by means of a short-range complex potential
  with two dimensionless couplings $c_0$ and $\lambda_0$,
  which respectively represent the short-range attraction and annihilation
  as explained in the main text.
  For $c_0 = 0$ (upper pannel), i.e. no short-range attraction,
  the bound state survives regardless of the magnitude of
  the annihilation amplitude.
  For $c_0 = 0.8$ (middle pannel), which corresponds to moderate short-range
  attraction (close to generating an excited bound state),
  the $Y_{c\bar c}$ state will become unbound
  for $\lambda_0 \geq 0.44$.
  For $c_0 = 1.6$ (lower panel) the short-range attraction is strong enough
  as to generate a fundamental and excited state.
  The fundamental state (red line) crosses the threshold
  for $\lambda_0 \geq 0.81$,
  while the excited state (green line)
  binds regardless of short-range annihilation.
}
\label{Ycc-width}
\end{figure}


The previous idea can be applied to discuss the effects of
annihilation in the $Y_{c \bar c}(5045)$ baryonium.
If we allow the short-range delta-shell potential to be complex
the imaginary piece will represent $\Lambda_{c1} \bar{\Sigma}_c$ annihilation.
As before we define the reduced couplings
\begin{eqnarray}
  c_0 + i\,\lambda_0 = -\frac{2 \mu_Y C_0}{4 \pi R_c}
  \, ,
\end{eqnarray}
where $c_0$ and $\lambda_0$ are dimensionless and real,
with $\lambda_0$ the coupling that models annihilation.
We show the movement of the $Y_{c \bar c}(5045)$ pole in Fig.~\ref{Ycc-width}
for $c_0 = 0$ (no short-range attraction),
$c_0 = 0.8$ (a moderate short-range attraction which is on the verge
of generating an excited state) and $c_0 = 1.6$ (strong short-range
attraction generating a fundamental and excited state).
For $c_0 = 0$ the molecular pole survives regardless of the strength of
the short-range annihilation amplitude.
Though not shown in Fig.~\ref{Ycc-width}, for short-range repulsion the
movement of the pole is pretty similar to the $c_0 = 0$ case.
For $c_0 = 0.8$ and $\lambda_0 \geq 0.44$ the pole moves above the threshold
but is still located in the first Riemann sheet,
which means that in practice it is as if it vanishes.
Conversely if we increase the short-range attraction to the point of
having an excited and fundamental state (lower panel in Fig.~\ref{Ycc-width}),
the fundamental state vanishes after $\lambda_0$ reaches a critical value
while the excited state survives for arbitrary values of $\lambda_0$,
eventually becoming the new fundamental state.
It is worth mentioning that a more realistic representation
should take into account that the range of the annihilation mechanism
and the short-range attraction/repulsion are not the same:
annihilation is expected to happen at distances considerably smaller
than the $R_c = 1\,{\rm fm}$ cut-off we have used for the delta-shell.
If anything this implies that our calculation will overestimate
the annihilation width.
Vector meson exchange is likely to be attractive
for the $\Lambda_c \bar{\Sigma}_c$ baryon-antibaryon system,
but this depends on the spin.
For $S=0$ both $\omega$ and $\rho$ exchange are attractive,
while for $S=1$ $\omega$ is attractive and $\rho$ repulsive.
Ignoring $\rho$ exchange and including OPE, $\sigma$ and $\omega$ exchange,
calculations in the OBE model with a monopolar
form factor of $\Lambda = 0.9-1.2\,{\rm GeV}$
indicate the existence of a shallow bound state at $0.1-0.2\,{\rm MeV}$
and a deep bound state at $15-130\,{\rm MeV}$.
For $S=0$ this suggest that we are probably
in the third scenario of Fig.~\ref{Ycc-width},
in which we have a resilient shallow and a fragile deep state.
Meanwhile for $S=1$ it depends on the interplay
between $\rho$ and $\omega$ exchange.
Thus the survival of the $Y_{c\bar c}(5045)$ to annihilation seems to be likely,
though it ultimately depends on how reliable is the phenomenology
of short-range dynamics we are using here~\footnote{
At this point it is interesting to compare the $\Lambda_{c1} \bar{\Sigma}_c$
baryonium with the more well-known nucleon-antinucleon baryonium.
The nucleon-antinucleon interaction is very attractive
at short distances because of $\omega$-exchange,
from which there should be a rich spectrum of
molecular states.
But annihilation probably prevents this spectrum
to be realized~\cite{Myhrer:1976by,Myhrer:1976ka}
(maybe with exceptions if the $p \bar{p}$ enhancement observed
by the BES collaboration~\cite{Bai:2003sw} turns out to be a baryonium).
The reason why the $\Lambda_{c1} \bar{\Sigma}$ baryonium
is more resilient to annihilation is the unnaturally
large range of the binding mechanism.
The natural scale at which annihilation happens is about a few ${\rm GeV}$,
which is to be compared with an effective pion mass of $20\,{\rm MeV}$.
In contrast for nucleon-antinucleon, which probably binds because vector
meson exchange, we have to compare the annihilation scale
with the $\omega$/$\rho$ meson mass.
Notice that if the $\Lambda_{c1} \bar{\Sigma}_c$ baryonium were
to be bound because of short-range attraction then
the conclusions would be different.
}.


To summarize,
the $\Lambda_{c1} \Sigma_c$ and $\Lambda_{c1} \bar{\Sigma}_c$ systems
can exchange an S-wave pion almost on the mass shell giving rise
to a potential with an unusual long range.
This Coulomb-like OPE extends to distances large enough as to have
at least a shallow S-wave state.
The spectrum might very well include additional states owing to
the short-range interaction between the baryons.
The baryon-antibaryon states appear both in the $S=0,1$ configurations,
where the G-parity is required to be $(-1)^{S+L+1}$
for Coulomb-like OPE to be attractive.
For the baryon-baryon states OPE is attractive for $S=0$, for which
the long-range piece of the potential is identical to the previous case.
They should bind even if the unknown short-range interaction
is strongly repulsive.
The shallow nature of these states begs the question of whether 
they survive once we take into account the finite width of
the heavy baryons. The answer is probably yes, but a deeper
theoretical analysis than the one presented here
would be very welcome.
In addition the baryon-antibaryon states can annihilate, i.e. decay
into charmonium, heavy meson-antimeson pairs and so on. Preliminary
calculations indicate that the survival $\Lambda_{c1} \bar{\Sigma}_c$
states is possible but depends on the short-range dynamics.
Finally we stress that
the mechanism binding the $\Lambda_{c1} \Sigma_c$ and
$\Lambda_{c1} \bar{\Sigma}_c$ extends to other systems
composed of two opposite parity hadrons
with the same spin and a mass difference similar to that of
a pseudo-Goldstone boson.

\section*{Acknowledgments}

This work is partly supported by the National Natural Science Foundation
of China under Grants No. 11375024,  No.11522539, No. 11735003
and the Fundamental Research Funds for the Central Universities.


%

\end{document}